# Vibration-Rotational Dynamics of Low-Mass Exotic Baryons

David Akers*
*Lockheed Martin Corporation, Dept. 6F2P, Bldg. 660, Mail Zone 6620,*
*1011 Lockheed Way, Palmdale, CA 93599*
*Email address: David.Akers@lmco.com

A Goldstone boson with a mass of 35 MeV is derived from the dibaryon masses in the experimental data of B. Tatischeff *et al.* The existence of quantized vibrations is utilized to explain narrow nucleon resonances below the pion-nucleon threshold. The quantum of 35 MeV is found between nucleon states with like spins and parities. Evidence of the 35 MeV quantum is in agreement with the mass operator method of the Fil'kov model for low-mass baryons and with the "light" pion concept of the Walcher particle-hole model. Azimov's idea of a low-lying baryon octet is revived with evidence from earlier experiments. The idea of rotational excitations is applied to the band structure of low-mass baryons in this paper.

## I. INTRODUCTION

The existence of quantized vibrations is derived from the experimental dibaryon masses in this paper. We utilized the idea of quantized vibrations and rotational dynamics to explain the spectra of low-mass exotic baryons. The existence of light quanta has its beginnings in the earlier work of Mac Gregor and in the Goldstone boson model. In this paper, we show that the vibration-rotational model is able to explain numerous experimental data. The existence of a 35 MeV Goldstone boson is given in Section II. In Section III, we show how the existence of light quanta with masses 35 and 70 MeV are able to account for the low-mass baryons. We present a comparison of the 35 MeV quantum with the mass separations between states with like spins and parities as found in the Fil'kov model. We revive the idea of Azimov and suggest the existence of more than one set of low-lying baryon octets. We postulate also the existence of band structure for low-mass baryons. Finally, we present concluding remarks in Section IV.

## II. LOW-MASS BARYONS

In recent years, experiments reported narrow excited states of the nucleon below the $\pi$ threshold [1-4]. Peaks were observed at 1004, 1044, and 1094 MeV with a width of about 5 MeV [1]. The first experiments involved $pp \to p\,\pi^+ N^*$. The second experiments found three peaks at 966, 985, and 1003 MeV [2]. These experiments involved the reaction $pd \to ppX_1$, where $X_1 = N^* \to \gamma\,N$, and measured the missing mass $X_1$. In addition, low-mass baryons can be derived from the masses of the dibaryons [5]. On the other hand, experiments of electron scattering on hydrogen/deuterium targets have indicated null signals [6, 7]. L'vov and Workman have argued that these narrow nucleon resonances are not observed in real Compton scattering experiments and must be excluded [8]. Thus, the existence of these reported narrow states has been met with disbelief.

However, the null results of electro-production experiments do not mean that the hadro-production experiments are faulty. In fact, we would argue that exotic baryons have unique production mechanisms. Particle physics is not like nuclear physics in the sense that large compound nuclei are created via several production channels. Exotic particle resonances, especially with narrow and supernarrow widths, indicate unique production mechanisms. By narrow, we mean widths with $\Gamma \sim 5\text{-}15$ MeV for baryons, and by supernarrow, we mean widths that are $\Gamma \ll 1$ keV for dibaryons [9]. The observations of narrow states below the pion-nucleon threshold have revived an old idea by Azimov of a low-lying exotic baryon octet [10]. From the numerous observations of baryon resonances [1-4], from the spectrum of dibaryons [5] and from the earlier observations of low-lying baryons [11], we suggest that there is more than one set of low-lying exotic baryon octets and that the original idea of Azimov is correct. The basis for this statement is that the spectrum of low-mass baryons can be explained by the existence of a 35 MeV quantum of vibration and by rotational dynamics as first suggested by Mac Gregor [11]. Therefore, the experimental data [1-5] are real, and there are unique production mechanisms for the generation of exotic baryons.

The existence of a 35 MeV Goldstone boson was proposed in an earlier paper [12] and was also derived from a modified QCD Langrangian, where magnetic monopoles were proposed as Goldstone bosons [13]. The 35 MeV quantum is also found in the Nambu empirical mass formula [14]. In addition, Walcher suggested the idea of a Goldstone boson or quantum to explain the dibaryon states [15]. Earlier, Walcher had proposed a particle-hole model from nuclear physics to explain these states [15]. Walcher proposed a light "pion" to account for equidistance between masses from all known experiments in Ref. [1-3]. An average mass difference of $\delta m = 21.2 \pm 2.6$ MeV was calculated as a fit to the mass series 940, 966, 985, 1003, $m_4$, 1044, $m_6$, and 1094 MeV, where $m_4 = 1023$ and $m_6 = 1069$ MeV. Thus, a light quantum of excitation, $m_{light\,\pi} = 2m_q = 21$ MeV, was proposed by Walcher [15].

It is noted that none of the isospins, spins and parities of the reported nucleon states have been measured because of the difficulties with weak signals and measuring angular distributions [1-5]. Hence, reaching an inference for equidistance between these nucleon masses must be met with caution since there may be different quantum numbers among the states.

From a study of the experimental dibaryon masses [5], Walcher determined that the dibaryon masses were equidistant and that there existed an excitation quantum of 36.7 MeV as calculated from his fit to the spectrum. There was no attempt to correct for the effects of the binding energies of the dibaryons. However, Mac Gregor has shown that typical binding energies are about 3-4% in many baryons [16]. Therefore, the equidistant hypothesis for the dibaryons should be nearly exact, as shown in Fig. 3 of Ref. [15].

In Fig. 1 of the present work, we plot the dibaryon masses (in MeV) as a function of the number n = 0, 1, 2, … of the excitation quantum. An intercept is calculated at about 1830 MeV, and a quantum is derived to be 35 MeV for the slope of the line in Fig. 1. The dibaryon masses fit a straight line to the formula:

$$m = 1830 \text{ MeV} + n \cdot 35 \text{ MeV}, \qquad (1)$$

where n = 0, 1, 2, … It is recognized immediately that the 35 MeV quantum is part of the Nambu empirical mass formula [14]. The 35 MeV quantum, along with the entire Nambu series, was later derived in a modified QCD Lagrangian [13]. The 35 MeV quantum is one-half of the famous 70 MeV quantum proposed by Mac Gregor [16, 17]. The Nambu mass series includes 0, 35, 70, 105, 140, 175, 210, … MeV. We suggest that the hν = 35 MeV quantum is a radial vibration of the nucleon. Synonymous terms would be a radial oscillation, an excitation, and the monopole (E0) or breathing mode [18].

## III. VIBRATION-ROTATIONAL DYNAMICS

The earliest known research into a study of the rotational spectra of baryons and mesons can be attributed to Mac Gregor from the 1970s [19, 20]. In the work of Mac Gregor, a constituent-quark (CQ) model was proposed that featured baryons and mesons as rotational states, which suggested a correlation between increasing L values and increasing energies in these resonances. The baryon resonances followed the expected interval rule of L(L + 1) for rotational spectra, according to the expression

$$E(L) = E_0 + (\hbar)^2 L(L+1)/2I,$$

$$E_{rot} = (\hbar)^2/2I, \qquad (2)$$

where $E_0$ is the bandhead energy, $E_{rot}$ is the rotational energy, and $I$ is the moment of inertia. Recently, the rotational spectra of mesons and baryons were studied in Ref. [21]. The experimental particle data are derived from the *Review of Particle Properties* [22]. Moreover, the rotational spectra of exotic baryons were also studied in Ref. [23].

The nature of the quantum with mass 35 MeV is discussed in Ref. [23], and it involves the quantized vibrations of the nucleon. Evidence is shown for nucleon vibrations in steps of 70 and 140 MeV [16, 21, 24]. Many models of baryons utilize a Hamiltonian, which involves harmonic oscillator potentials as the confining force for short-range distances [25]. The 3-dimensional harmonic oscillator potentials result in quantized vibrations with the energy $E = (N + 3/2)h\nu$ [25]. It is the general belief that the vibrations and rotations will mix in the baryon states. This result comes from the molecular model of baryons [26]. In the molecular model, the vibration and rotational modes are mixed in the Hamiltonian calculations. However, in the chiral soliton model, the quantized vibrations are relatively independent of the rotations in baryons [27]. From the experimental data [21], the spectra of nucleons indicate linear relationships between the rotational energy and the angular momentum in agreement with Eq. (2). Thus, the experimental data support the conclusion of Diakonov and Petrov [27].

In Fig. 2, we show the rotational spectra of the light nucleons from 939 to approximately 1800 MeV. Clearly, a pattern of 70 MeV separations is seen between nucleons. For the low-mass baryons, we identify the experimental evidence for each nucleon not listed in the *Review of Particle Properties* [22]. The existence of the N(1359)S peak is given in πN scattering experiments [28-32]. The N(1359)S has a width Γ < 67 MeV and can be explained in the framework of the rotational model by Mac Gregor [33]. The N(1413)P is actually the measured nucleon state by Hirose *et al*. [32] and is listed under the Roper resonance N(1440) by the PDG [22]. The dashed lines in Fig. 2 indicate possible nucleon states with 70 MeV separations from those nucleons represented by the solid lines. The N(1219)S is an unobserved S-state baryon bandhead in Eq.(43) of Ref. [11], but the mass is noted from the dibaryon spectrum as the state 2155 – 939 = 1216 MeV in Fig. 18 of Ref. [5]. The corresponding rotational partner of the N(1219)S is the N(1300)P. The N(1300)P is found in several experiments [34-39]. The N(1079) and N(1149)S may have some evidence (see Fig. 17 in Ref. [5] and Table 3 in Ref. [4]). The evidence for the N(1094)P state is found in Ref. [4]. Evidence for the N(1017) is noted in Ref. [40]. The estimated N(1169)D is based upon a rotational energy of 27 MeV, according to Eq. (2). It is noted that there is a N(1171) in Table 1 of Ref. [4], which is very close to the predicted N(1169)D. Thus, we can see that there are several low-mass baryon states, which are not listed in the *Review of Particle Properties* (RPP).

We note that the chiral soliton model does not predict these low-mass baryon states [27]. Moreover, lattice QCD physics does not account for these narrow width baryons. In fact, the only models which allow for these baryons are the particle-hole model of Walcher [15], the mass operator method in the Fil'kov model [9], and the vibration-rotational model of the present paper. It remains to be seen whether the Goldstone boson exchange (GBE) model of Glozman will become a predictive tool to account for the low-mass exotic baryons, while the GBE model has been successful to some degree with the baryons (see Fig. 4 in Ref. [41]).

If we consider the spectrum of Fig. 2 to be the first level for evidence of quantized vibrations in units of 70 MeV, then Fig. 3 may be considered the second level with an even smaller unit of 35 MeV. In Fig. 3, the scale is expanded from 904 to 1254 MeV, showing quantized vibrations in units of 35 MeV as indicated by the arrow. The following nucleons are found in both Figs. 2 and 3: N(1219)S, N(1149)S, N(1094)P, N(1079), N(1017), N(1009), and N(939). References to the experimental data for these nucleons have already been presented and will not be repeated. Evidence for N(1114) and N(1184) comes from the dibaryon mass spectrum (see Fig. 17 in Ref. [5] and references therein). The evidence for N(986)P can be found in the results of Fil'kov, Kashevarov, and Konobeevski [42]. The N(950) state is noted in the work by Tatischeff [43]. Thus, there is clear evidence of a quantum with a mass 35 MeV in Fig. 3.

As a side note, we mention the fact that the N(939) is chosen as a P-state by convention since baryon number is conserved [44]. In Figs. 2 and 3, it would seem that N(939) should be an S-state or ground state, which presents a problem for the vibration-rotational model. We offer several possible suggestions to explain this problem: 1) the vibration-rotational model is simply in error, 2) there is indeed a low-lying nucleon with a mass ~ 941 MeV which is an S-state, 3) there are virtual states below the ground state N(939), 4) there is a weak interaction between the 35 MeV Goldstone boson and the nucleon N(939) which violates parity, and 5) the chiral soliton model is correct that all these low-lying baryons have positive parity as noted in Fig. 2 of Ref.[27].

In fact, the chiral soliton model predicts rotational excitations, which form a sequence of bands (see Fig. 2 in Ref. [27]). If the spectrum of Fig. 2 can be considered the first level for evidence of quantized vibrations and Fig. 3 may be considered the second level, then there must exist an even finer third level which forms a band structure of the low-mass baryons, according to the chiral solition model. In the present vibration-rotational model and in the Fil'kov model, these low-lying baryons should have both positive and negative parity states unlike the chiral soliton model. In Fig. 4, we propose a possible band structure of the low-mass baryons. This proposed band structure is based upon evidence from the Fil'kov model [9] and is shown in Table I. In Table I, we have listed the predicted nucleon states according to the mass operator method of the Fil'kov model. In addition, we have listed the S- and P-states from the vibration-rotational model in the same table. The existence of a 35 MeV quantum spacing is shown between states with like spins and parities. The first N* state comes from the Fil'kov model. In Table I, we show evidence for a N(950)P, which is below the first excited N* state at the predicted energy of 963.4 MeV. Thus, we suggest that there may exist an S-state below the N(950)P, and we estimate a N(941)S as shown in Fig. 4. The exact spectrum, which Fig. 4 is only representative, remains to be determined from experiments.

*Note added in proof:* At the time of the writing of this paper, the author received a preprint from B. Tatischeff on the experimental status of low-mass exotic baryons [45]. From Table I in Ref. [45], we note evidence for the predicted states N(1235), N(1370) and N(1580)P in Fig. 2 of the present paper. In Fig. 3, there is evidence for N(1200) and N(1235) from Table I in Ref. [45]. In addition, there is evidence for N(1104) and N(1113), which we show as solid lines in Fig. 4, and these are also found from Table I in Ref. [45].

As a final note, we mention again the idea of Azimov that there should exist low-lying states forming a baryon octet [10]. We suggest also that there should exist a baryon decuplet in the low-mass region. Earlier evidence of a Δ(1149)S was presented in the work of Mac Gregor [11]. Mac Gregor cited the experiment of P. E. Argen *et al.* (see p. 1305 and note 137 in Ref. [11]). The experiment of P. E. Argen *et al*. may be found in Ref. [46]. In addition, we would expect the existence of a Δ(1052)P or Δ(1079)P to fill in the table of Ref. [47].

## IV. CONCLUSION

In this paper, a Goldstone boson with a mass of 35 MeV was derived from the dibaryon masses. The existence of the 35 MeV quantum was derived earlier in 1994 from a modified QCD Lagrangian and forms the basis for the Nambu empirical mass formula. The existence of quantized vibrations and rotational dynamics was utilized to explain narrow nucleon resonances below the pion-nucleon threshold. The quantum of 35 MeV was found to exist between nucleon states with like spins and parities as shown in the Fil'kov model. Diakonov's and Petrov's idea of band structure was applied to the low-mass baryons. We expect that the concept of band structure will be further applied to higher nucleon states.

## ACKNOWLEDGEMENT

The author wishes to thank Malcolm Mac Gregor, retired from the University of California's Lawrence Livermore National Laboratory, for his encouragement to pursue the CQ Model, and he wishes to thank Paolo Palazzi of CERN for his interest in the work. The author thanks Boris Tatischeff for his preprint in Ref. [45], Lev Fil'kov for his challenge to explain some low-mass baryons, and Thomas Walcher for e-mail correspondence.

# FIGURE CAPTIONS

**Fig. 1.**  The experimental dibaryon masses against the number of excitation quanta with unit mass 35 MeV.  Experimental data are from Fig. 18 in Ref. [5], where the average mass difference is computed to be 34.8 MeV for the quantum (*cf.* Fig. 8 in Ref. [24]).

**Fig. 2.**  The rotational spectra of the light nucleons are shown.  The solid lines represent baryons supported by experimental evidence, and the dashed lines are predicted states.

**Fig. 3.**  The vibration-rotational spectra of low-mass baryons are shown with a scale of 35 MeV.  Note the nucleons are equally spaced in units of 35 MeV.

**Fig. 4.**  The band structure of the low-mass baryons is proposed for the represented nucleons.

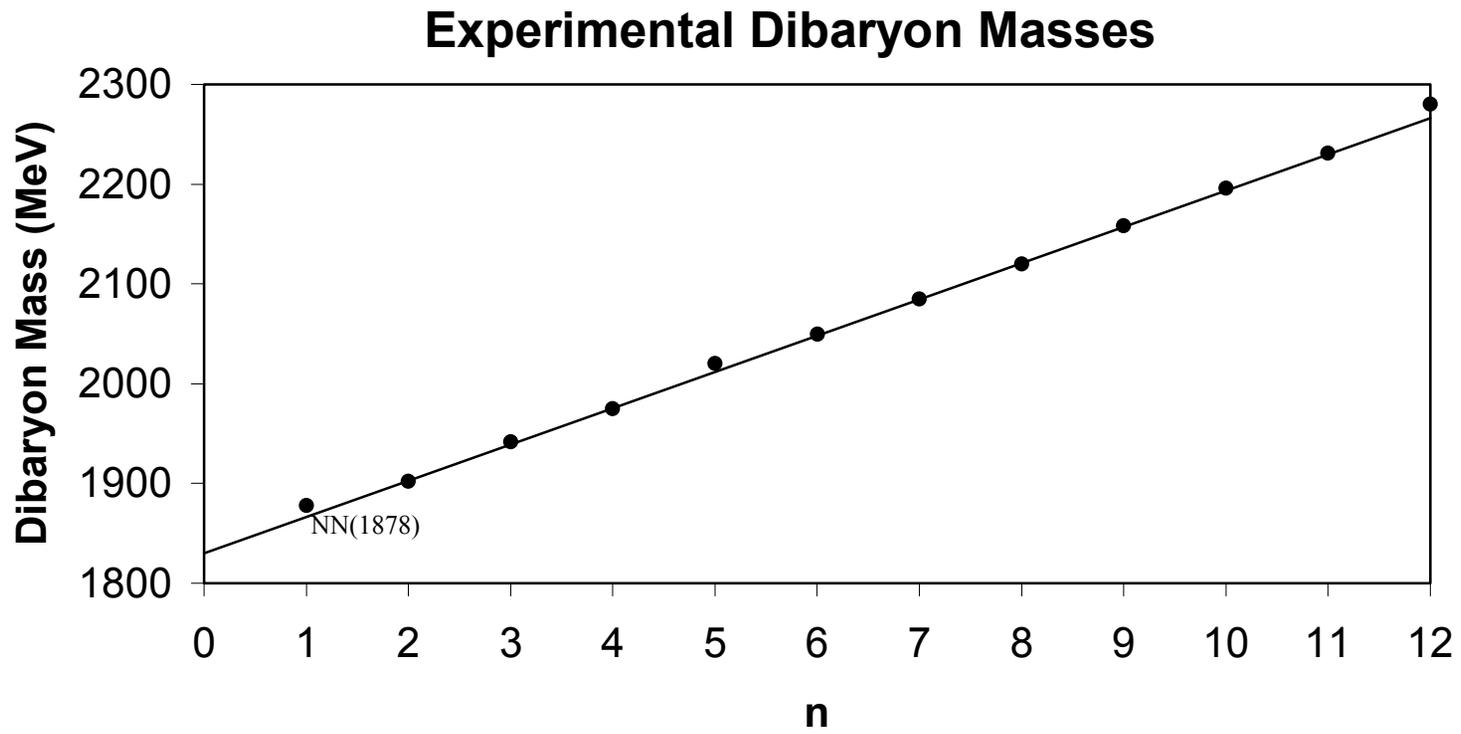

Fig. 1.

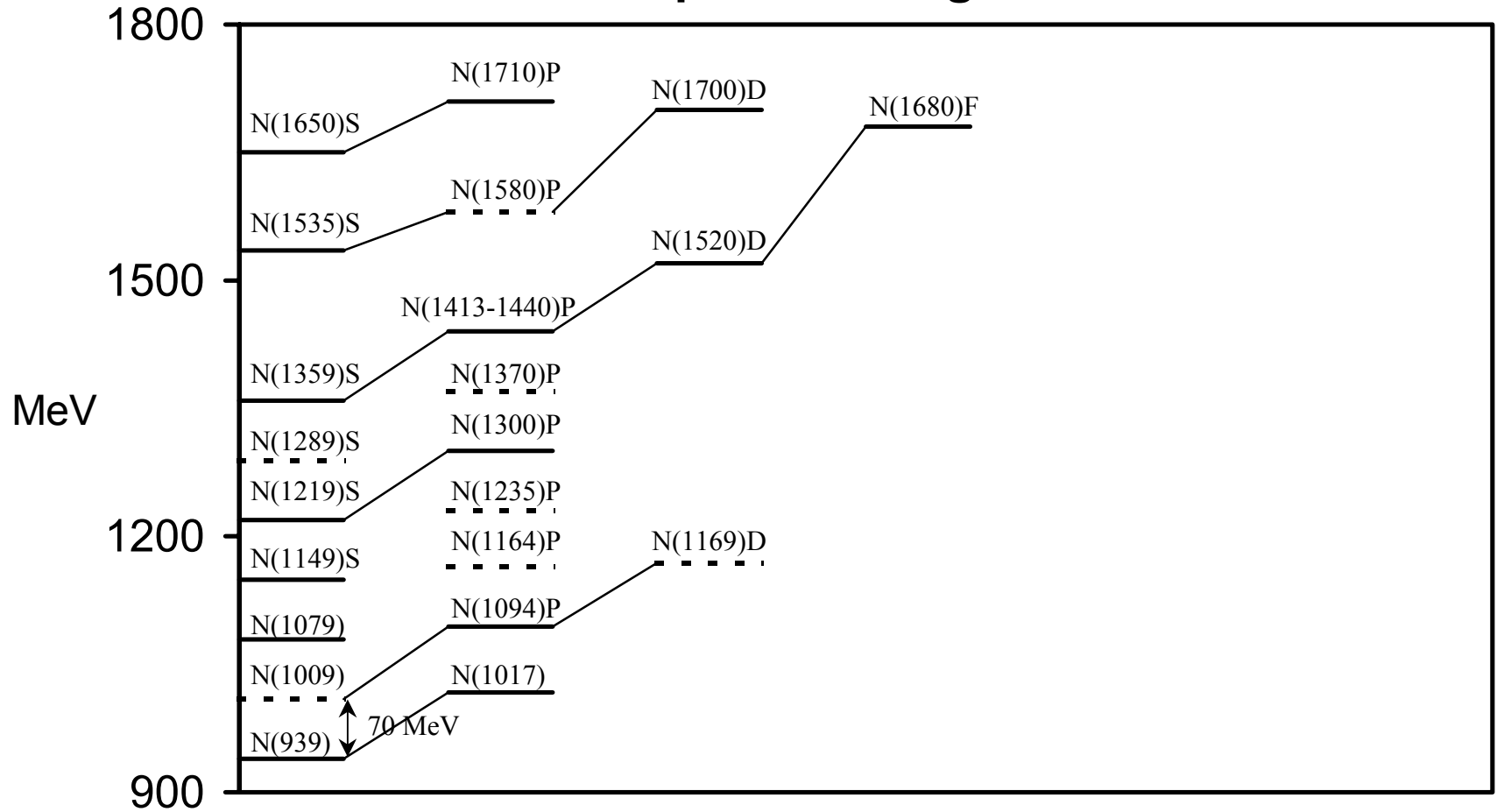

**Fig. 2.**

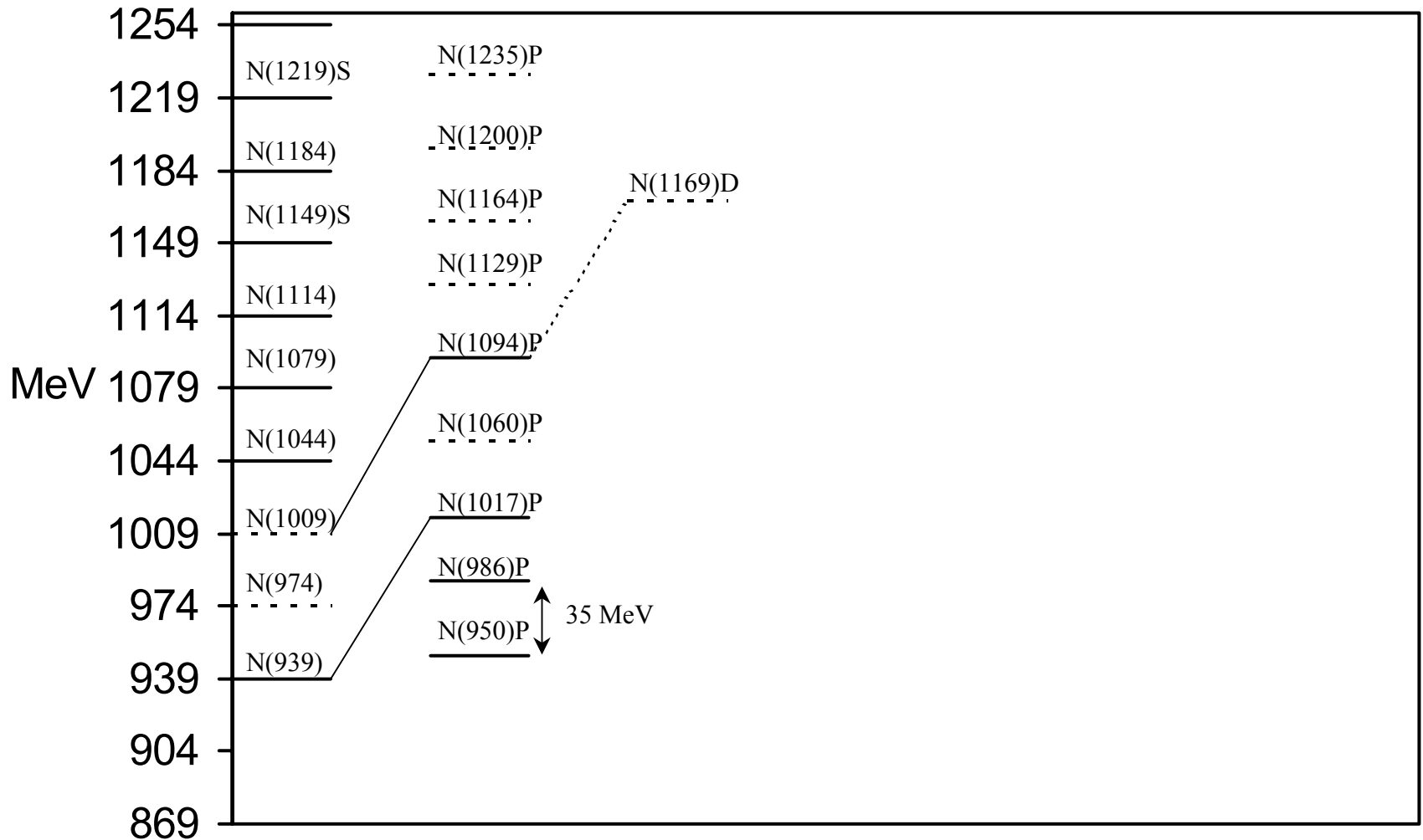

Fig. 3.

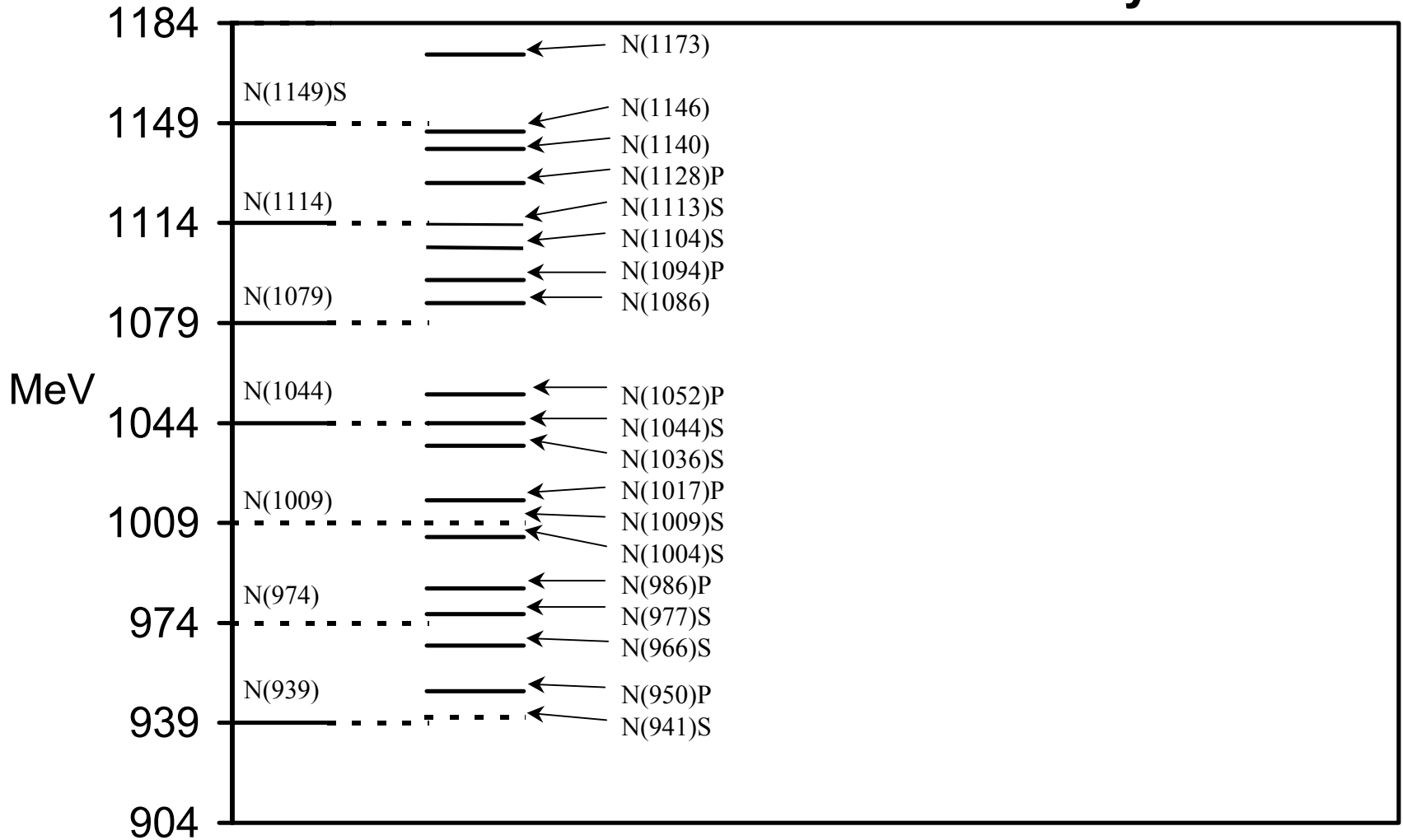

**Fig. 4.**

Table I. The masses and $J^P$ of the exotic baryon resonances from Fil'kov [9]. A comparison of the Fil'kov model is made with the existence of a 35 MeV quantum spacing between states with like spins and parities.

| State | S-state Nucleons | P-State Nucleons | N* | $J^P$ | Fil'kov Model (MeV) | Experiment (MeV) | Experimental Reference |
|---|---|---|---|---|---|---|---|
| P |  | N(950) |  |  |  | 950 | [43] |
| S | N(974) |  | 1 | $½^-$ | 963.4 | 966 ± 2, 977 | [42], [5] |
| P |  | N(985) | 2 | $½^+$ | 987 | 986 ± 2 | [42] |
| S | N(1009) |  | 3 | $½^-$ | 1010 | 1004 ± 2 | [1, 42] |
| P |  | N(1020) | 4 | $½^+$ | 1033 | 1017 ± 6 | [40] |
| S | N(1044) |  | 5 | $½^-$ | 1056 | 1044 | [1] |
| P |  | N(1055) | 6 | $½^+$ | 1079 | ? |  |
| S | N(1079) |  | 7 | $½^-$ |  | 1079 | [5] |